\documentclass[11pt]{article}
\usepackage{fleqn,espcrc1}
\usepackage{graphicx}
\usepackage{here}
\newcommand{\be}{\begin{equation}}
\newcommand{\ee}{\end{equation}}
\newcommand{\bea}{\begin{eqnarray}}
\newcommand{\eea}{\end{eqnarray}}

\def\beq{\begin{equation}}
\def\eeq{\end{equation}}
\def\bea{\begin{eqnarray}}
\def\eea{\end{eqnarray}}
\def\bq{\begin{quote}}
\def\eq{\end{quote}}

\def\nnb{\nonumber}
\def\ga{\left(}
\def\dr{\right)}
\def\aga{\left\{}
\def\adr{\right\}}

\def\rar{\rightarrow}
\def\lrar{\Longrightarrow}

\def\nnb{\nonumber}
\def\la{\langle}
\def\ra{\rangle}
\def\nin{\noindent}
\def\ba{\begin{array}}
\def\ea{\end{array}}

\def\als{\alpha_s}

\def\g2{ \la\alpha_s G^2 \ra}
\def\g3{g^3f_{abc}\la G^aG^bG^c \ra}
\def\g4{\la\als^2G^4\ra}

\newcommand{\AmS}{{\protect\the\textfont2
  A\kern-.1667em\lower.5ex\hbox{M}\kern-.125emS}}

\title{\vskip-6cm{\baselineskip14pt
{\normalsize\hfill PM/00--14}\\
{\normalsize\hfill KEK--TH/00--689}\\
{\normalsize\hfill Revised (August 2000)}\\
\vskip3cm
NEW QCD--ESTIMATE OF THE KAON PENGUIN MATRIX ELEMENTS
AND $\epsilon'/\epsilon$}
\vskip1.5cm}
\author{
Stephan Narison\address{Laboratoire de Physique Math\'ematique et Th\'eorique,
Universit\'e de Montpellier II
Place Eug\`ene Bataillon,
34095 - Montpellier Cedex 05, France \\ and \\ KEK Theory group, 1-1-Oho,
Tsukuba-city, Ibaraki 305, Japan.\\
E-mail:
narison@lpm.univ-montp2.fr}
}

\begin{document}

\maketitle
\vskip1.5cm
\begin{abstract}
\nin
Firstly, we use the recent ALEPH/OPAL data on the $V-A$ spectral functions
for fixing the continuum threshold
with
which the first and second Weinberg sum rules should be satisfied in the
chiral limit. Then, we predict the values
of the low-energy constants $m_{\pi^+}-m_{\pi^0}$, ${ L}_{10}$, and test
the values of
the electroweak kaon penguin matrix elements
$\la {\cal Q}^{3/2}_{8,7}\ra_{2\pi}$ obtained from DMO--like
sum rules. Secondly, we use the data on the $\tau$-total hadronic width
$R_{\tau,V/A}$ for extracting $\la
{\cal Q}_8^{3/2}\ra_{2\pi}$, in the $\overline{MS}$--scheme, and propose
some new
sum rules for $\la{\cal Q}_7^{3/2}\ra_{2\pi}$ in the chiral limit, where
the latter require
more accurate data for the spectral functions near the
$\tau$-mass. Thirdly, we analyze the effects to the matrix element $\la
{\cal Q}_6^{1/2}\ra_{2\pi}$, of the
$S_2\equiv(\bar uu+\bar dd)$  component of the $I=0$ scalar meson, with its
parameters
fixed from QCD spectral sum rules. Our results should stimulate a
further attention on the r\^ole of the (expected large) gluonium component of the
$I=0$ scalar meson and of the
associated operator in the $K\rar\pi\pi$ amplitude.
Finally, using our previous determinations, we deduce, in the Standard Model (SM), the conservative
upper bound  for the CP-violating
ratio: $\epsilon'/\epsilon \leq (22\pm 9) 10^{-4}$ , which is in
agreement with the present measurements.

\end{abstract}
\vfill\eject
\setcounter{page}{1}
\pagestyle{plain}
\textwidth 15.5cm
\textheight 23.5cm
\topmargin -3.cm
\oddsidemargin +0.2cm
\evensidemargin -1.0cm
\section{INTRODUCTION AND GENERALITIES}
\nin
$CP$--violation is one of the most important weak interactions phenomena in
particle
physics \cite{RAFAEL,STEIN}, where in their presence, stable (under strong
interactions) $K^0(\bar sd)$ and
$\bar K^0(\bar ds)$ particles with definite strangeness eigenvalues $\pm$1
become unstable. Therefore, the
decay of a long-lived kaon
\footnote{$\vert K_L\ra$ and $\vert K_S\ra$ are very close to the
$CP$-eigenstates
$\vert K^0_1\ra\equiv\frac{1}{\sqrt{2}}\ga
\vert K^0\ra-\vert
\bar{K}^0\ra\dr
$ and $\vert K_2^0\ra\equiv \frac{1}{\sqrt{2}}\ga \vert K^0\ra+\vert
\bar{K}^0\ra\dr
$ with $CP \vert K^0_1\ra=+\vert K^0_1\ra$ and $CP \vert K^0_2\ra=-\vert
K^0_2\ra$. Their lifetime are
$\tau_L\simeq 5.2\times 10^{-8}$ sec $\approx 15.51$ m $\approx 5.8\times
10^{2}\tau_S$. Their mass-difference
is $\Delta m\equiv M_L-M_S=(3.522\pm 0.016)\times 10^{-12}$ MeV.}
into a two-pion final state is an evidence for $CP$-violation. The first
observation of such a transition to the
$\pi^+\pi^-$ mode, was discovered 36 years ago by \cite{CRONIN} from
$K^0$-$\bar{K}^0$ mixing. Since then, the transition into $\pi^0\pi^0$ and
the phases of the
ratio of the amplitudes:
\beq
\eta_{+-}\equiv \frac{A\ga K_L\rar\pi^+\pi^-\dr}{A\ga
K_S\rar\pi^+\pi^-\dr}~,~~~~~~~~~
\eta_{00}\equiv \frac{A\ga K_L\rar\pi^0\pi^0\dr}{A\ga K_S\rar\pi^0\pi^0\dr}~,
\eeq
 have been also
observed. For a
phenomenological analysis, it is convenient to work with quantities where
the final pion
states are in a definite isospin. Then, one introduces the indirect
(through $K^0$-$\bar{K}^0$ mixing)
$CP$--violation parameter $\epsilon$, and the quantity $\omega$ governing
the so-called $\Delta=1/2$ rule
(enhancement of the $I=0$ over the $I=2$ transitions):
\beq
\epsilon\equiv\frac{A[K_L\rar(\pi\pi)_{I=0}]}
{A[K_S\rar(\pi\pi)_{I=0}]}~,~~~~~~~~~~~~
\omega\equiv\frac{A[K_S\rar(\pi\pi)_{I=2}]} {A[K_S\rar(\pi\pi)_{I=0}]}~,
\eeq
and the direct (in the amplitude) $CP-$violation parameter:
\beq
\epsilon'\equiv\frac{1}{\sqrt{2}}\aga\frac{A[K_L\rar(\pi\pi)_{I=2}]}
{A[K_S\rar(\pi\pi)_{I=0}]}-\epsilon \times
\omega\adr~.
\eeq
In terms of these quantities, one can express the measured $\eta_{+-}$ and
$\eta_{00}$ quantities as:
\beq
\eta_{+-}=\epsilon+\frac{\epsilon'}{1+\omega/\sqrt{2}}~,~~~~~~~~~~~~~~~~~
\eta_{00}=\epsilon-\frac{2\epsilon'}{1-\sqrt{2}\omega}~,
\eeq
from which one can deduce the experimental value \cite{PDG}:
\beq
\epsilon\simeq  (2.280\pm 0.013)\times 10^{-3} e^{i(43.5\pm 0.1)^0}~. \eeq
To proceed further, one introduces the isospin amplitude:
\beq
A\ga K^0\rar(\pi\pi)_{I}\dr =iA_Ie^{i\delta_I},~~~~A\ga
\bar{K}^0\rar(\pi\pi)_{I}\dr
=-iA^*_Ie^{i\delta_I}~,
\eeq
and makes use of the mass-difference $\Delta m\equiv m_L-m_S$, between
$K_S$ and $K_L$,  and of $M_{12}$ (off-diagonal dispersive part of the
$K^0$-$\bar{K}^0$
complex mass matrix: Re $M_{12}\simeq \Delta m/2$).
Therefore, one can write, within a good approximation \cite{RAFAEL}:
\beq\label{def1}
\epsilon\simeq \frac{1}{\sqrt{2}}e^{i\frac{\pi}{4}}\ga \frac{{\rm
Im}M_{12}}{\Delta m}+
\frac{{\rm
Im}A_0}{{\rm Re}A_0}\dr~,~~~~~~~~~~~~~~~~~~~~\omega\approx
e^{i(\delta_2-\delta_0)}\frac{{\rm Re}A_2} {{\rm Re}A_0}~.
\eeq
and:
\bea\label{def2} \epsilon'\simeq
\frac{1}{\sqrt{2}}e^{i(\delta_2-\delta_0+\frac{\pi}{2})}\frac{{\rm
Re}A_2} {{\rm Re}A_0}\ga \frac{{\rm Im}A_2} {{\rm
Re}A_2}-\frac{{\rm Im}A_0}{{\rm Re}A_0}\dr
\eea
Experimentally \cite{PDG}, $\delta_2-\delta_0\simeq  -(42\pm 4)^0$, and
then, using the data on
$\Gamma(K_S\rar\pi^+\pi^-)/\Gamma(K_S\rar\pi^0\pi^0)$, one can deduce:
\bea
\omega_{exp}&\simeq& \ga\frac{1}{22}\dr e^{-i(42\pm 4)^0}~,
\eea
while the recent measurement reported by the kTeV and NA48 experiments
\cite{KTEV} on the direct $CP-$violation
ratio is \footnote{If one includes the preliminary NA48 result ${\rm
Re}({\epsilon'}/{\epsilon})\simeq (12.2\pm 2.9(stat.)\pm 4(syst.))\times
10^{-4}$ from the 98 data sample, the
preliminary new experimental world average becomes ${\rm
Re}({\epsilon'}/{\epsilon})\simeq (19.3\pm 2.4)\times
10^{-4}$ \cite{TATI}.}:
\beq\label{exp}
{\rm Re}\ga\frac{\epsilon'}{\epsilon}\dr_{exp}\simeq (21.4\pm 4.0)\times
10^{-4}~.
\eeq
It is fair to say that a simultaneous explanation of these two previous
experimental numbers remains a challenge
for the present theoretical predictions within the Standard Model (SM)
\cite{RAFAEL,STEIN,BURAS,OTHER,MARTI,FAB}.
\section{THEORY OF ${\epsilon'}/{\epsilon}$}
\nin
In the SM, it is customary to study the $\Delta S=1$ process from the weak
hamiltonian:
\beq
{\cal
H}_{eff}=\frac{G_F}{\sqrt{2}}V_{ud}V^*_{us}\sum_{i=1}^{10}{C_i(\mu)Q_i(\mu)}~,
\eeq
where $C_i(\mu)$ are perturbative Wilson Coefficients known including
complete NLO QCD corrections \cite{BURAS},
which read in the notations of \cite{BURAS}:
\beq
C_i(\mu)\equiv z_i(\mu)-\frac{V_{td}V^*_{ts}}{V_{ud}V^*_{us}}y_i(\mu)~.
\eeq
where $V_{ij}$ are elements of the CKM-matrix; $Q_i(\mu)$ are
non-perturbative hadronic matrix elements
which need to be estimated from different non-perturbative methods of QCD
(chiral perturbation theory, lattice,
QCD spectral sum rules,...). In the choice of basis of \cite{BURAS}, the
dominant contributions come from the
four-quark operators which are classified as:
\begin{itemize}
\item {\bf Current-Current:}
\bea
{\cal Q}_1&\equiv& \ga \bar s_\alpha u_\beta\dr_{V-A}\ga \bar
u_\beta
d_\alpha\dr_{V-A}~,~~~~~~~~~~~~~~~~~~{\cal
Q}_2\equiv \ga \bar s u\dr_{V-A}\ga
\bar  u d\dr_{V-A}
\eea
\item{\bf QCD-penguins:}
\bea
{\cal Q}_3&\equiv& \ga \bar s d\dr_{V-A}\sum_{u,d,s}\ga \bar
\psi\psi\dr_{V-A}~,~~~~~~~~~~~~~~~~~~{\cal Q}_4\equiv
\ga \bar s_\alpha
d_\beta\dr_{V-A}\sum_{u,d,s}\ga \bar
\psi_\beta\psi_\alpha\dr_{V-A}~,\nnb\\
{\cal Q}_5&\equiv& \ga \bar s d\dr_{V-A}\sum_{u,d,s}\ga \bar
\psi\psi\dr_{V+A}~,~~~~~~~~~~~~~~~~~~{\cal Q}_6\equiv
\ga \bar s_\alpha
d_\beta\dr_{V-A}\sum_{u,d,s}\ga \bar
\psi_\beta\psi_\alpha\dr_{V+A}~.
\eea
\item{\bf Electroweak-penguins:}
\bea
{\cal Q}_7&\equiv& \frac{3}{2}\ga \bar s d\dr_{V-A}\sum_{u,d,s}e_\psi\ga \bar
\psi\psi\dr_{V+A}~,~~~~~~~~~~~~{\cal Q}_8\equiv
\frac{3}{2}\ga \bar s_\alpha
d_\beta\dr_{V-A}\sum_{u,d,s}e_\psi\ga \bar
\psi_\beta\psi_\alpha\dr_{V+A}~,\nnb\\
{\cal Q}_9&\equiv& \frac{3}{2}\ga \bar s d\dr_{V-A}\sum_{u,d,s}e_\psi\ga \bar
\psi\psi\dr_{V-A}~,~~~~~~~~~~~{\cal Q}_{10}\equiv
\frac{3}{2}\ga \bar s_\alpha
d_\beta\dr_{V-A}\sum_{u,d,s}e_\psi\ga \bar
\psi_\beta\psi_\alpha\dr_{V-A},\nnb\\
\eea
\end{itemize}
where $\alpha,\beta$ are colour indices; $e_\psi$ denotes the electric charges
\footnote{Though apparently suppressed, the effect of the electroweak
penguins are enhanced
by $1/\omega$ as we shall see later on in Eq. (\ref{ampli}).}
reflecting the electroweak nature of ${\cal Q}_{7,...10}$, while $V-(+)A \equiv
\ga 1-(+)\gamma_5\dr\gamma_\mu$. Using an OPE of the amplitudes, one obtains:
\beq
\frac{\epsilon'}{\epsilon} \simeq {\rm Im}\lambda_t\Big{[}
P^{(1/2)}-P^{(3/2)}\Big{]}e^{i\Phi}
\eeq
where $\Phi\equiv \Phi_{\epsilon'}-\Phi_{\epsilon}\approx 0$ {[}see Eqs.
(\ref{def1}) and (\ref{def2})];
$\lambda_t\equiv V_{td}V_{ts}^*$ can be expressed in terms of the CKM
matrix elements as ($\delta$ being the CKM
phase)
\cite{BURAS,ALI}:
\beq
{\rm Im} \lambda_t\approx \vert V_{ub}\vert\vert
V_{cb}\vert{\sin}{\delta}\simeq (1.33\pm
0.14)\times 10^{-4}~,
\eeq
from $B$-decays and $\epsilon$.
The QCD quantities $P^{(I)}$ read:
\bea\label{ampli}
P^{(1/2)}&=&\frac{G_F\vert \omega\vert}{2\vert \epsilon\vert{\rm Re}
A_0}\sum_iC_i(\mu)
\la (\pi\pi)_{I=0}|{\cal Q}_i|K^0\ra_0\ga 1-\Omega_{IB}\dr~,\nnb\\
P^{(3/2)}&=&\frac{G_F}{2\vert \epsilon\vert{\rm Re} A_2}\sum_iC_i(\mu)
\la (\pi\pi)_{I=2}|{\cal Q}_i|K^0\ra_2~.
\eea
 $\Omega_{IB}
\simeq (0.16\pm 0.03)$   quantifies the $SU(2)$--isospin breaking effect,
which includes the one of the
$\pi^0$--$\eta$  mixing \cite{PP}, and which reduces the usual value of
$(0.25\pm 0.08)$ \cite{BURAS} due to
$\eta'$--$\eta$ mixing.  It is also expected that the QCD- and
electroweak-penguin operators:
\bea\label{penguine}
{\cal Q}^{3/2}_8
\approx B^{3/2}_8/m_s^2+{\cal O}(1/N_c)~,~~~~~~~~~~~~~~~~~~~
{\cal Q}^{1/2}_6 \approx B^{1/2}_6/m_s^2+{\cal O}(1/N_c)~,
\eea
give the dominant contributions to the ratio
$\epsilon'/\epsilon$ \cite{VSZ}; $B$ are the bag factors which are expected
to be 1 in the large $N_c$-limit.
Therefore, a simplified approximate but very informative expression of the
theoretical predictions can be
derived \cite{BURAS}:
\bea\label{epsilon}
 \frac{\epsilon'}{\epsilon}\approx 13~{\rm Im}\lambda_t
\ga\frac{110}{\overline{m}_s(2)~[{\rm MeV}]}
\dr^2\times\Bigg{[}B^{1/2}_6\ga 1-\Omega_{IB}\dr -0.4B^{3/2}_8\ga
\frac{m_t}{165~{\rm GeV}}\dr
\Bigg{]}\ga \frac{\Lambda^{(4)}_{\overline MS}}{340~{\rm MeV}}\dr~,
\eea
\begin{table}
\setlength{\tabcolsep}{1.pc}
\begin{center}
\caption{ Penguin $B$--parameters for the $\Delta S=1$ process from
different approaches at $\mu=2$ GeV.
We use the value $m_s(2)=(119\pm 12)$ MeV from \cite{SNL}, and predictions
based on dispersion relations
\cite{DONO,ENJL} have been rescaled according to it. We also use for our
results $f_\pi=92.4$ MeV
\cite{PDG}, but we give in the text their $m_s$ and $f_\pi$ dependences.
Results without any comments on the scheme have been
obtained in the $\overline{MS}- NDR-$scheme. However, at the present accuracy, one cannot
differentiate these results from the 
ones of $\overline{MS}- HV-$scheme.}
\begin{tabular}[h]{ccccc}
\hline
& & &&\\
{\bf METHODS}&${\boldmath B^{1/2}_6}$&$B^{3/2}_8$&$B^{3/2}_7$&{\bf COMMENTS}\\
&&&&\\
\hline
&&&&\\
Lattice \cite{MARTI,ROME,GUPTA}&$0.6\sim 0.8$ &$0.7\sim 1.1$ & $0.5\sim
0.8$&Huge NLO\\
&unreliable&&&at matching\cite{KILCUP}\\
&&&&\\
Large $N_c$ \cite{HAMBYE}&$0.7\sim 1.3$&$0.4\sim 0.7$&$-0.10\sim
0.04$&${\cal O}(p^0/N_c,~p^2)$\\
&&&&scheme?\\
&$1.5\sim 1.7$&$-$&$-$&${\cal O}(p^2/N_c)$; $m_q=0$\\
&&&&scheme?\\
{\bf Models}&&&\\
Chiral QM \cite{FAB}&$1.2\sim 1.7$&$\sim 0.9$&$\approx
B^{3/2}_8$&$\mu=.8$ GeV \\
&&&&rel. with $\overline{MS}$ ?\\
ENJL+IVB \cite{PRADES}&$2.5\pm 0.4$&$1.4\pm 0.2$&$0.8\pm 0.1$&$NLO$ in $1/N_c$\\
&&&&$m_q=0$ \\
L$\sigma$-model \cite{SIGMAL}&$\sim 2$&$\sim 1.2$&$-$&Not unique\\
&&&&$\mu\approx 1$ GeV; scheme ?\\
NL $\sigma$-model \cite{SIGMAN}&$1.6\sim 3.0$&$0.7\sim
0.9$&$-$&$M_\sigma$: free; $SU(3)_F$ trunc.
\\ &&&&$\mu\approx 1$ GeV; scheme ?\\
&\\
{\bf Dispersive}\\
Large $N_c$+ LMD&$-$&$-$&0.9&$NLO$ in $1/N_c$, \\
+LSD--match.\cite{ENJL}&&&strong $\mu$-dep.&\\
&&&& \\
DMO-like SR \cite{DONO}&$-$&$1.6\pm 0.4$&$0.8\pm 0.2$&$m_q=0$\\
&&huge NLO&&Strong $s$,~ $\mu$--dep.\\ &&&\\
FSI \cite{PALLANTE}&$1.4\pm 0.3$&$0.7\pm 0.2$&$-$&Debate for fixing \\
&&&&the Slope \cite{BURAS2}\\
&&&\\
{\bf This work}&&&&\\
DMO-like SR:&--&$2.2\pm 1.5$&$0.7\pm 0.2$&$m_q=0$\\
\cite{DONO} revisited&&inaccurate&&Strong $s$,~ $\mu$--dep.\\
&\\
$\tau$-like SR&$-$&$-$&inaccurate&$t_c$--changes\\
&&&&\\
$R^{V-A}_\tau$&$-$&$1.7\pm 0.4$&$-$&$m_q=0$\\
&&&&\\
$S_2\equiv (\bar uu+\bar dd)$&$1.0\pm 0.4$&$-$&$-$&$\overline{MS}-$scheme\\
from QSSR&$\leq 1.5\pm 0.4$&&&$m_s(2)\geq$ 90 MeV\\
&&&\\
\hline
\end{tabular}
\end{center}
\end{table}

\nin
where the average value $\hat B_K= 0.80\pm 0.15$ of the $\Delta S=2$
process has been used. This value
includes the conservative value $0.58\pm 0.22$ from Laplace sum rules
\cite{SNBK}.
The values of the top quark mass and the QCD scale
$\Lambda^{(4)}_{\overline MS}$ \cite{PDG,BETHKE} are under a quite good
control and have small effects. A recent
review of the light quark mass determinations \cite{SNL} also indicates
that the strange quark mass is also under
control and a low value advocated in the previous literature to explain the
present data on
$\epsilon'/\epsilon$  is unlikely due the
lower bound constraints from the positivity of the QCD spectral function or
from the positivity of the $m^2$
corrections to the GMOR PCAC relation. For a consistency with the approach
used in this paper, we shall use the
average value of the light quark masses from QCD spectral sum rules(QSSR),
$e^+e^-$ and $\tau$-decays given in
\cite{SNL}:
\beq\label{ms}
\overline{m}_s(2)\simeq (119\pm 12)~{\rm MeV}~,~~~~~
\overline{m}_d(2)\simeq (6.3\pm 0.8)~{\rm MeV}~,~~~~~
\overline{m}_u(2)\simeq (3.5\pm 0.4)~{\rm MeV}~.
\eeq
Using the previous experimental values, one can deduce the constraint in
\cite{SNL} updated:
\beq\label{const}
{\cal B}_{68}\equiv B^{1/2}_6-0.48B^{3/2}_8\simeq 1.73\pm 0.50~({\rm resp.}
\geq 1.0\sim 1.2),
\eeq
if one uses the value of $m_s$ in Eq. (\ref{ms}) (resp. the lower bound of
$(90\sim 100)$ MeV
reported in
\cite{SNL}). This result
shows a possible violation of more than 2$\sigma$ for the leading
$1/N_c$ vacuum saturation prediction $ \approx 0.52$ corresponding to
$B_6^{1/2}\approx
B_8^{3/2}\approx 1$. Consulting the available predictions reviewed in
\cite{BURAS}, which we will
summarize and update in Table 1, one can notice that  the values of the
$B$--parameters have
large errors. One can also see that results from QCD first principles
(lattice and $1/N_c$)
fail to explain the data, which however can be accomodated by various
QCD-like models. We
shall come back to this discussion when we shall compare our results with
presently available predictions.
It is, therefore, clear that the present estimate of the four-quark
operators, and in particular the estimates
of the dominant penguin ones
given previously in Eq. (\ref{penguine}), need to be reinvestigated. Due to
the complex structures and large size
of these operators,
they should be difficult to extract unambiguously from different approaches.
In this paper, we present alternative theoretical approaches based as well
on first principles of QCD
($\tau$--decay data, analyticity), for
predicting the size of the QCD-- and electroweak--penguin operators given
in Eq. (\ref{penguine}).
In performing this analysis, we shall also encounter the electroweak
penguin operator:
\beq
{\cal Q}^{3/2}_7\
\approx B^{3/2}_7/m_s^2+{\cal O}(1/N_c)~,
\eeq
and some other low-energy constants ($m_{\pi^\pm}-m_{\pi^0}, ~ L_{10}$)
though not directly relevant to
$\epsilon'/\epsilon$.
\section{TESTS OF THE ``SACROSANTE"  WEINBERG AND DMO SUM RULES IN THE CHIRAL LIMIT}
\subsection{Notations}
\nin
Before estimating these condensates, we shall test the procedure used in
\cite{DONO} by analyzing
the classics DMO-- and Weinberg--like sum rules \cite{DMO,WEINBERG}. This
analysis will also allow us to fix
the cut-off parameter
$t_c$ until which the data on $V-A$ spectral functions from ALEPH/OPAL
\cite{ALEPH,OPAL} are known.
We shall be concerned here with the two-point correlator:
\beq
\Pi^{\mu\nu}_{LR}(q)\equiv i\int d^4 x~ e^{iqx}\la 0|{\cal T} J^\mu_L(x)\ga
J^\nu_R(0)\dr^{\dagger}|0\ra=-(g^{\mu\nu}q^2-q^\mu q^\nu)\Pi_{LR}(q^2)~,
\eeq
built from the left-- and right--handed components of the local weak current:
\beq
J^\mu_{L}=\bar u\gamma^\mu(1-\gamma_5)d,~~~~~~~J^\mu_{R}=\bar
u\gamma^\mu(1+\gamma_5)d~,
\eeq
and/or using isospin rotation relating the neutral and charged weak currents:
\beq
\rho_V-\rho_A\equiv \frac{1}{2\pi}{\rm Im}\Pi_{LR}\equiv\frac{1}{4\pi^2}\ga
v-a\dr~.
\eeq
The first term is the notation in \cite{DONO}, while the last one is the
notation in \cite{ALEPH,OPAL}.
\subsection{The sum rules}
\nin
The ``sacrosante" DMO and Weinberg sum rules read in the chiral limit
\footnote{Systematic analysis of the breaking of these sum rules by light
quark masses \cite{FNR} and
condensates \cite{SNWEIN,SVZ} within the context of QCD have been done
earlier.}:
\bea
{\cal S}_{0}&\equiv&\int_0^{\infty} ds ~\frac{1}{2\pi}{\rm Im}\Pi_{LR} =
f^2_\pi~,\nnb\\
{\cal S}_{1}&\equiv&\int_0^{\infty} ds ~s~\frac{1}{2\pi}{\rm Im}\Pi_{LR} =
0~,\nnb\\
{\cal S}_{-1}&\equiv&\int_0^{\infty} \frac{ds}{s}~ \frac{1}{2\pi}{\rm
Im}\Pi_{LR} = -4L_{10}~,\nnb\\
{\cal S}_{em}&\equiv&\int_0^{\infty} ds~\ga s~\log\frac{s}{\mu^2}\dr
\frac{1}{2\pi}{\rm Im}\Pi_{LR} =
-\frac{4\pi}{3\alpha}f^2_\pi \ga m^2_{\pi^\pm}-m^2_{\pi^0}\dr~,
\eea
where $f_\pi\vert_{exp}=(92.4\pm 0.26)$ MeV is the experimental pion decay
constant which should be used
here as we shall use data from $\tau$-decays involving physical pions;
$m_{\pi^\pm}-m_{\pi^0}\vert_{exp}\simeq 4.5936(5)$ MeV; $L_{10}\equiv
f_\pi^2{\la r_\pi^2\ra}/{3}-F_A$
{[}$\la r_\pi^2\ra=(0.439\pm 0.008)fm^2$ is the mean pion radius and
$F_A=0.0058\pm 0.0008$ is the
axial-vector pion form factor for $\pi\rar e\nu\gamma${]} is one the
low-energy constants of the effective
chiral Lagrangian
\cite{RAFAEL}. In order to exploit these sum rules using the ALEPH/OPAL
\cite{ALEPH,OPAL} data from the hadronic tau--decays, we shall work with
their Finite
Energy Sum Rule (FESR) versions (see e.g. \cite{FNR,BNP} for such a
derivation). In the chiral
limit  ($m_q=0$ and
$\la
\bar uu\ra=\la
\bar dd\ra =\la \bar ss\ra$), this is equivalent to truncate the LHS at
$t_c$ until which the data are available, while the RHS of the integral
remains valid to leading order
in the 1/$t_c$ expansion in
the chiral limit, as the breaking of
these sum rules by higher dimension $D=6$ condensates in the chiral limit
which is of the order of
$1/t_c^3$ is numerically negligible \cite{SNWEIN}.
\subsection{Matching between the low and high-energy regions}
\nin
In order to fix the $t_c$ values which separate the low and high energy
parts of the spectral functions, we
require that the 2nd Weinberg sum rule (WSR) ${\cal S}_1$ should be
satisfied by the present data. 
\begin{figure}[H]
\begin{center}
\caption{ FESR version of the 2nd Weinberg sum rule versus $t_c$ in GeV$^2$
using the ALEPH/OPAL data of the spectral functions. Only the central values are
shown.}
\end{center}
\end{figure}
\nin
As shown
in Fig. 1, this is obtained for two
values of $t_c$ \footnote{One can compare the two solutions with the
$t_c$--stability region around 2 GeV$^2$ in the
QCD spectral sum rules analysis (see e.g. Chapter 6 of \cite{SNB}).}:
\beq\label{weinberg}
t_c\simeq (1.4\sim 1.5)~{\rm GeV}^2~~~~~{\rm and}~~~~~
t_c\simeq (2.4\sim 2.6)~{\rm GeV}^2.
\eeq
Though the 2nd value is interesting from the point of view of the QCD
perturbative calculations (better convergence of the QCD series), its exact
value is strongly affected by the
inaccuracy of the data near the
$\tau$--mass (with the low values of the ALEPH/OPAL data points, the 2nd
Weinberg sum rule is only
satisfied at the former value of $t_c$).\\
After having these $t_c$ solutions, we can improve the constraints by
requiring that the 1st Weinberg sum rule ${\cal
S}_0$ reproduces the experimental value of
$f_\pi$ \footnote{Though we are working here in the chiral limit, the data
are obtained for physical pions,
such that the corresponding value of $f_\pi$ should also correspond to the
experimental one.}
within an accuracy 2-times the experimental error. 
This condition allows to fix
$t_c$ in a very narrow margin due to the sensitivity of the result on the
changes of $t_c$ values
\footnote{For the
second set of
$t_c$-values in Eq. \ref{weinberg}, one obtains a slightly lower value:
$f_\pi=(84.1\pm 4.4)$ MeV.}
\beq\label{tc}
t_c=(1.475\pm 0.015)~{\rm GeV}^2~,
\eeq
\section{
LOW-ENERGY CONSTANTS $L_{10}$, $m_{\pi^\pm}-m_{\pi^0}$ AND $f_\pi$ IN THE CHIRAL LIMIT
}
\nin
Using the previous value of $t_c$ into the ${\cal S}_{-1}$ sum rule, we deduce:
\beq\label{l10}
L_{10}\simeq -(6.26\pm 0.04)\times 10^{-3}~,
\eeq
which agrees quite well with more involved analysis including chiral
symmetry breakings \cite{STERN,OPAL},
and with the one using a lowest meson dominance (LMD) of the spectral
integral \cite{ENJL}. \\
Analogously, one obtains from the ${\cal S}_{em}$ sum rule:
\beq\label{deltam}
\Delta m_\pi\equiv m_{\pi^\pm}-m_{\pi^0}\simeq (4.84\pm 0.21)~{\rm MeV}~.
\eeq
This result is 1$\sigma$ higher than the data $4.5936(5)$ MeV, but agrees
within the errors with the more detailed
analysis from $\tau$--decays \cite{PECCEI,OPAL} and with the LMD result of
about 5 MeV \cite{ENJL}. We have checked
that moving the subtraction point $\mu$ from 2 to 4 GeV slightly decreases
the value of
$\Delta m_\pi$ by $3.7\%$ which is relatively weak, as expected. Indeed, in
the chiral limit, the $\mu$
dependence does not appear (to leading order in $a_s$) in the RHS of the ${\cal
S}_{em}$ sum rule, and then, it looks
natural to choose:
\beq
\mu^2=t_c~,
\eeq
as $t_c$ is the only external scale in the analysis. At this scale the
result increases
slightly by 2.5\%. One can also notice that the prediction for $\Delta m$
is more stable when one changes the
value of $t_c=\mu^2$. Therefore, the final predictions from the value of
$t_c$ in Eq. (\ref{tc}) fixed from the 1st
and 2nd Weinberg sum rules are:
\beq\label{pred}
\Delta m \simeq (4.96\pm 0.22)~{\rm MeV}~,~~~~~~~~~~L_{10}\simeq -(6.42\pm
0.04)\times 10^{-3}~,
\eeq
which we consider as our "best" predictions. \\
For some more conservative results, we also give the predictions obtained from
the second $t_c$--value given in Eq. (\ref{weinberg}). In this way, one
obtains:
\beq
f_\pi=(87\pm 4)~{\rm MeV}~,~~~~~~\Delta m \simeq (3.4\pm 0.3)~{\rm
MeV}~,~~~~~~L_{10}\simeq -(5.91\pm
0.08)\times 10^{-3}~,
\eeq
where one can notice that the results are systematically lower
than the ones obtained in Eq. (\ref{pred}) from the first $t_c$--value
given previously, which may disfavour
a posteriori the second choice of $t_c$-values, though we do not have a
strong argument favouring one with respect
to the other \footnote{Approach based on
$1/N_c$ expansion and a saturation of the spectral function by the lowest
state within a narrow width
approximation (NWA) favours the former value of $t_c$ given in Eq.
(\ref{tc}) \cite{ENJL2}.}. 
Therefore, we take as a conservative value the largest range
spanned by the two sets of results,
namely:
\beq
f_\pi=(86.8\pm 7.1)~{\rm MeV}~,~~~~~~~~\Delta m \simeq (4.1\pm 0.9)~{\rm
MeV}~,~~~~~~~~~~L_{10}\simeq -(5.8\pm
0.2)\times 10^{-3}~,
\eeq
which we found to be quite satisfactory in the chiral limit.
The previous tests are very useful, as they will allow us to jauge the
confidence level of the next predictions.
\section{SOFT PION AND KAON REDUCTIONS OF $\la (\pi\pi)_{I=2}|{\cal
Q}^{3/2}_{7,8}|K^0\ra$}
\nin
An interesting approach combining pion and kaon reductions in the chiral
limit with dispersion relation
techniques have been proposed recently \cite{DONO} \footnote{A similar
approach based on large $N_c$ and a
lowest meson dominance (LMD) of the spectral functions is also
done in \cite{ENJL}.} in order to estimate   the matrix element:
\beq
\la {\cal Q}^{3/2}_{7,8}\ra_{2\pi}\equiv \la (\pi\pi)_{I=2}|{\cal
Q}^{3/2}_{7,8}|K^0\ra~.
\eeq
In the chiral limit $m_{u,d,s}\sim m^2_\pi\simeq m^2_K=0$, one can use soft
pion and kaon techniques in order
to relate the previous amplitude to the four-quark vacuum condensates:
\bea\label{soft}
\la {\cal Q}^{3/2}_{7}\ra_{2\pi}&\simeq &-\frac{4}{f_\pi^3}\la
{\cal O}^{3/2}_7\ra~,\nnb\\
\la {\cal Q}^{3/2}_{8}\ra_{2\pi}&\simeq
&-\frac{4}{f_\pi^3}\aga\frac{1}{3}\la {\cal O}^{3/2}_7\ra+\frac{1}{2}\la
{\cal O}^{3/2}_8\ra\adr~,
\eea
where we use the shorthand notations: $\la 0|{\cal O}^{3/2}_{7,8}|0\ra\equiv
\la {\cal O}^{3/2}_{7,8}\ra $, and $f_\pi=(92.42\pm 0.26)$ MeV \footnote{In
the chiral limit $f_\pi$ would be
about 84 MeV. However, it is not clear to us what value of $f_\pi$ should
be used here, so we shall leave it as
a free parameter which the reader can fix at his convenience.}. Here:
\bea
{\cal O}^{3/2}_7\equiv {\cal
O}^{3/2}_1\cite{DONO}&=&\sum_{u,d,s}
\bar\psi\gamma_\mu\frac{\tau_3}{2}\psi\bar\psi\gamma_\mu\frac{\tau_3}{2}\psi-
\bar\psi\gamma_\mu\gamma_5\frac{\tau_3}{2}\psi\bar\psi\gamma_\mu\gamma_5\frac{
\tau_3}{2}\psi~,\nnb\\
{\cal O}^{3/2}_8&=&\sum_{u,d,s}
\bar\psi\gamma_\mu\lambda_a\frac{\tau_3}{2}\psi\bar\psi\gamma_\mu\lambda_a\frac{
\tau_3}{2}\psi-
\bar\psi\gamma_\mu\gamma_5\lambda_a\frac{\tau_3}{2}\psi\bar\psi\gamma_\mu\gamma_
5\lambda_a\frac{\tau_3}{2}\psi~,
\eea
where $\tau_3$ and $\lambda_a$ are flavour and colour matrices.
Using further pion and kaon reductions in
the chiral limit, one can relate this matrix element to the
$B$-parameters:
\bea\label{soft2}
B^{3/2}_7(M^2_\tau)&\simeq& \frac{3}{4}\frac{\ga m_u+m_d\dr
}{m^2_\pi}\frac{\ga m_u+m_s\dr
}{m^2_K}\frac{1}{f_\pi}
\la {\cal Q}^{3/2}_{7}\ra_{2\pi}(M^2_\tau)\nnb\\
B^{3/2}_8(M^2_\tau)&\simeq& \frac{1}{4}\frac{\ga m_u+m_d\dr
}{m^2_\pi}\frac{\ga m_u+m_s\dr
}{m^2_K}\frac{1}{f_\pi}
\la {\cal Q}^{3/2}_{8}\ra_{2\pi}(M^2_\tau)
\eea
where all QCD quantities will be evaluated in the $\overline{MS}$-scheme
and at the scale $M_\tau$.
\section{ $\la (\pi\pi)_{I=2}|{\cal Q}^{3/2}_{7,8}|K^0\ra$ FROM DMO--LIKE
SUM RULES IN THE CHIRAL LIMIT}
\nin
In a
previous paper \cite{DONO}, the vacuum condensates $
\la {\cal O}^{3/2}_{7,8}\ra $ which are related to the weak matrix elements
$\la (\pi\pi)_{I=2}|{\cal
Q}^{3/2}_{7,8}|K^0\ra$ through the soft pion and kaon reduction techniques
(see previous section) have been extracted
using Das-Mathur-Okubo(DMO)-- and Weinberg--like sum rules based on the
difference of the vector and axial-vector
spectral functions $\rho_{V,A}$  of the
$I=1$ component of the neutral current:
\bea\label{dispersive}
2\pi\la \alpha_s {\cal O}^{3/2}_8\ra(\mu^2)&=&\int_0^\infty ds
~s^2\frac{\mu^2}{s+\mu^2}\ga
\rho_V-\rho_A\dr(s)~,\nnb\\
\frac{16\pi^2}{3}\la {\cal O}^{3/2}_7\ra(\mu^2)&=&\int_0^\infty ds~
s^2\log\ga\frac{s+\mu^2}{s}\dr\ga
\rho_V-\rho_A\dr(s)~,
\eea
where $\mu$ is the subtraction point.
Due to the quadratic divergence of the integrand, the previous
sum rules are expected to be sensitive to the high energy tails of the
spectral functions where the
present ALEPH/OPAL data from $\tau$-decay \cite{ALEPH,OPAL} are inaccurate.
This inaccuracy can a
priori affect the estimate of the four-quark vacuum condensates. On the
other hand,
the explicit
$\mu$--dependence of the analysis can also induce another uncertainty.
En passant, we check below the effects of these two parameters $t_c$ and $\mu$.
After evaluating the spectral integrals, we obtain at $\mu$= 2 GeV and for
our previous values of $t_c$ in Eq. (\ref{tc}), the values (in units of
$10^{-3}$ GeV$^6$)
using the cut-off momentum scheme (c.o):
\beq\label{mom}
\alpha_s\la{\cal O}^{3/2}_{8}\ra_{c.o}\simeq -(0.69\pm 0.06)~,~~~~~~~~~~~~~~~
\la{\cal O}^{3/2}_{7}\ra_{c.o}\simeq -(0.11\pm 0.01)~,
\eeq
where the errors come mainly from the small changes of $t_c$--values. If
instead, we use the second set of values
of $t_c$ in Eq. (\ref{weinberg}), we obtain by setting $\mu$=2 GeV:
\beq\label{mom2}
\alpha_s\la{\cal O}^{3/2}_{8}\ra_{c.o}\simeq -(0.6\pm 0.3)~,~~~~~~~~~~~~~~~
\la{\cal O}^{3/2}_{7}\ra_{c.0}\simeq -(0.10\pm 0.03)~,
\eeq
which is consistent with the one in Eq. (\ref{mom}), but with larger errors
as expected. We have also checked that
both $\la{\cal O}^{3/2}_{8}\ra$ and $\la{\cal O}^{3/2}_{7}\ra$ increase in
absolute value when $\mu$ increases where
a stronger change is obtained for $\la{\cal O}^{3/2}_{7}\ra$, a feature
which has been already noticed in
\cite{ENJL}. In order to give a more conservative estimate, we consider as
our final value the largest range
spanned by our results from the two different sets of $t_c$--values. This
corresponds to the one in Eq. (\ref{mom2})
which is the less accurate prediction.
 We shall
use the relation between the momentum cut-off (c.o) and
$\overline{MS}$--schemes given in
\cite{DONO}:
\bea\label{rel}
\la{\cal O}^{3/2}_{7}\ra_{\overline{MS}}&\simeq& \la{\cal O}^{3/2}_{7}\ra_{c.o}
+\frac{3}{8}a_s\ga \frac{3}{2}+2d_s\dr\la{\cal O}^{3/2}_{8}\ra\nnb\\
\la{\cal O}^{3/2}_{8}\ra_{\overline{MS}}&\simeq& \ga 1-\frac{119}{24}a_s
\pm \ga\frac{119}{24}a_s\dr^2\dr\la{\cal O}^{3/2}_{8}\ra_{c.o}
-a_s\la{\cal O}^{3/2}_{7}\ra~,
\eea
where $d_s=-5/6$ (resp 1/6) in the so-called Na\"\i ive Dimensional Regularization NDR (resp.
t'Hooft-Veltmann HV) schemes \footnote{The two schemes differ by the treatment of the $\gamma_5$
matrix.};
$a_s\equiv
\alpha_s/\pi$. One can notice that the $a_s$ coefficient is large in the
2nd relation (50\%
correction) \cite{OKADA}, and the situation is worse because of the
relative minus sign
between the two contributions. Therefore, we have added a rough estimate of
the $a_s^2$ corrections based on the na\"\i ve growth of the PT series,
which here gives 50\%
corrections of the sum of the two first terms. For
a consistency of the whole approach, we shall use the value of
$\alpha_s$ obtained from
$\tau$--decay, which is \cite{ALEPH,OPAL}:
\beq
\alpha_s(M_\tau)|_{exp}=0.341\pm 0.05~\lrar\alpha_s(2~{\rm GeV})\simeq
0.321\pm 0.05~.
\eeq
Then, we deduce (in units of $10^{-4}$ GeV$^6$) at 2 GeV:
\beq\label{dono78}
\la{\cal O}^{3/2}_{7}\ra_{\overline{MS}}\simeq -(0.7\pm 0.2)~,~~~~~~~~~~~
\la{\cal O}^{3/2}_{8}\ra_{\overline{MS}}\simeq -(9.1\pm 6.4)~,
\eeq
where the large error in $\la{\cal O}^{3/2}_{8}\ra$ comes from the estimate
of the $a_s^2$ corrections
appearing in Eq. (\ref{rel}). In terms of the $B$ factor and with the
previous value of
the light quark masses in Eq. (\ref{ms}), this result, at $\mu=2$ GeV, can be
translated into:
\bea\label{b78}
B^{3/2}_{7}&\simeq& (0.7\pm 0.2)~\ga\frac{m_s(2)~[{\rm
MeV}]}{119}\dr^2\ga\frac{92.4}{f_\pi~[{\rm MeV}]}\dr^4~,\nnb\\
B^{3/2}_{8}&\simeq&(2.5\pm 1.3)~\ga\frac{m_s(2)~[{\rm
MeV}]}{119}\dr^2\ga\frac{92.4}{f_\pi~[{\rm
MeV}]}\dr^4~.
\eea
\begin{itemize}
\item
Our results in Eqs. (\ref{dono78}) compare quite well with the ones obtained by
\cite{DONO} in the $\overline{MS}$--scheme (in units of $10^{-4}$ GeV$^6$)
at 2 GeV:
\beq
\la{\cal O}^{3/2}_{8}\ra_{\overline{MS}}\simeq -(6.7\pm 0.9)~,~~~~~~~~~~~
\la{\cal O}^{3/2}_{7}\ra_{\overline{MS}}\simeq -(0.70\pm 0.10)~,
\eeq
using the same sum rules but presumably a slightly different method for the
uses of the data and for the choice
of the cut-off in the evaluation of the spectral integral.
\item Our errors in the evaluation of the spectral integrals, leading to
the values in Eqs.
(\ref{mom}) and (\ref{mom2}), are mainly due to the slight change of the
cut-off value $t_c$ \footnote{A slight
deviation from such a value affects notably previous predictions as the
$t_c$-stability of the results ($t_c\approx 2$ GeV$^2$) does not coincide
with the one required by the 2nd
Weinberg sum rules. At the stability point the predictions are about a
factor 3 higher than the one obtained
previously.}.
\item The error due to the passage into the ${\overline{MS}}$--scheme is
due mainly
to the truncation of the QCD series, and is important (50\%) for $\la{\cal
O}^{3/2}_{8}\ra$ and $B^{3/2}_8$,
which is the main source of errors in our estimate.
\item As noticed earlier, in the analysis of the pion mass-difference, it
looks more natural to do the
subtraction at $t_c$. We also found that moving the value of
$\mu$ can affects the value of $B^{3/2}_{7,8}$ .
\end{itemize}
For the above reasons, we expect that the results
given in
\cite{DONO} for $\la{\cal O}^{3/2}_{8}\ra$ though interesting are quite
fragile, while the errors quoted there
have been presumably underestimated. Therefore, we think that a
reconsideration of these results using
alternative methods are mandatory.
\section{$\la (\pi\pi)_{I=2}|{\cal Q}^{3/2}_{8}|K^0\ra$ FROM THE HADRONIC
TAU TOTAL DECAY RATES}
\nin
In the following, we shall
not introduce any new sum rule, but, instead, we shall exploit known
informations from the
total $\tau$--decay rate and available results from it, which have not the
previous drawbacks. The $V-A$ total
$\tau$--decay rate, for the $I=1$ hadronic component, can be deduced from
\cite{BNP} (hereafter referred as BNP),
and reads \footnote{Hereafter we shall work in the $\overline{MS}$--scheme.}:
\beq\label{rate}
{ R}_{\tau,V-A}=\frac{3}{2}\vert V_{ud}\vert
^2S_{EW}\sum_{D=2,4,...}{\delta^{(D)}_{V-A}}~.
\eeq
$\vert V_{ud}\vert =0.9753\pm 0.0006$ is the CKM-mixing angle, while
$S_{EW}=1.0194$ is the electroweak corrections
\cite{MARC}. In the following, we shall use the BNP results for ${\cal
R}_{\tau,V/A}$ in order to deduce ${
R}_{\tau,V-A}$:
\begin{itemize}
\item The chiral invariant
$D=2$ term due to a short distance tachyonic gluon mass \cite{ZAK,CNZ}
cancels in the $V-A$ combination.
 Therefore, the $D=2$ contributions come only from the quark mass terms:
\beq
M^2_\tau\delta^{(2)}_{V-A}\simeq 8\Big{[}
1+\frac{25}{3}a_s(M_\tau)\Big{]}m_u(M_\tau)m_d(M_\tau)~,
\eeq
as can be obtained from the first calculation \cite{FNR}, where
$a_s(M_\tau)\equiv \alpha_s/\pi(M_\tau)$ and $m_u(M_\tau)\simeq (3.5\pm 0.4)$
MeV,
$m_d(M_\tau)\simeq (6.3\pm 0.8)$ MeV \cite{SNL}  are respectively
the running coupling and  quark masses evaluated at the scale $M_\tau$.
\item The dimension-four condensate
contribution reads:
\beq
M^4_\tau\delta^{(4)}_{V-A}\simeq 32\pi^2\ga 1+\frac{9}{2}a_s^2\dr m^2_\pi
f^2_\pi+ {\cal O} \ga m^4_{u,d}\dr~,
\eeq
where we have used the $SU(2)$ relation $\la \bar uu\ra=\la \bar dd\ra$ and
the Gell-Mann-Oakes-Renner PCAC
relation:
\beq
(m_u+m_d)\la
\bar uu+\bar dd\ra=-2m^2_\pi f^2_\pi~.
\eeq
\item By inspecting the structure of the combination of dimension-six
condensates entering in
${ R}_{\tau,V/A}$ given by BNP \cite{BNP}, which are renormalizaton group
invariants, and using a $SU(2)$
isospin rotation which relates the charged and neutral (axial)--vector
currents, the $D=6$ contribution reads:
\beq\label{o8}
M^6_\tau\delta^{(6)}_{V-A}=-2\times 48\pi^4a_s\aga\ga 1+\frac{235}{48}a_s\pm
\ga\frac{235}{48} a_s\dr^2-\frac{\lambda^2}{M^2_\tau}\dr \la
{\cal O}^{3/2}_8\ra+a_s\la {\cal O}^{3/2}_7\ra\adr~,
\eeq
where the overall factor 2 in front expresses the different normalization
between the neutral isovector and
charged currents used respectively in \cite{DONO} and \cite{BNP}, whilst
all quantities are evaluated at the
scale $\mu=M_\tau$. The last two terms in the Wilson coefficients of $\la
{\cal O}^{3/2}_8\ra$ are new: the first term is an estimate of the NNLO term by
assuming a na{\"\i}ve geometric  growth of the $a_s$ series; the second one
is the effect of a tachyonic gluon
mass introduced in
\cite{CNZ}, which takes into account the resummation of the QCD asymptotic
series, with:
$a_s\lambda^2\simeq -0.06$ GeV$^2$
\footnote{This contribution may compete with the dimension-8 operators discussed
in \cite{DONO3}.}. Using
the values of
$\alpha_s(M_\tau)$ given previously, the corresponding QCD series behaves quite well as:
 \beq
{\rm Coefficient~ of}~\la{\cal O}^{3/2}_8\ra\simeq  1+(0.53\pm 0.08)\pm
0.28+0.18~,
\eeq
where the first error comes from the one of $\alpha_s$, while the second
one is due to the unknown
$a_s^2$--term, which introduces an uncertainty of 16\% for the whole
series. The last term is due to the
tachyonic gluon mass.
This leads to the numerical value:
\beq
M^6_\tau\delta^{(6)}_{V-A}\simeq -(1.015\pm 0.149)\times 10^3\aga (1.71\pm
0.29)\la
{\cal O}^{3/2}_8\ra+a_s\la {\cal O}^{3/2}_7\ra\adr~,
\eeq
\item If, one estimates the $D=8$ contribution using a vacuum saturation
assumption, the
relevant $V-A$ combination vanishes to leading order of the chiral symmetry
breaking terms. Instead, we
shall use the combined ALEPH/OPAL \cite{ALEPH,OPAL} fit for
$\delta^{(8)}_{V/A}$, and deduce:
\beq
\delta^{(8)}_{V-A}\vert_{exp}=- (1.58\pm 0.12)\times 10^{-2}~.
\eeq
\end{itemize} We shall also use the  combined ALEPH/OPAL data for ${
R}_{\tau,V/A}$, in order to obtain:
\beq
{ R}_{\tau,V-A}\vert_{exp}= (5.0\pm 1.7)\times 10^{-2},~~~~~~~~~~~
\eeq
Using the previous informations into the expression of the rate given in
Eq. (\ref{rate}), one can
deduce:
\beq
\delta^{(6)}_{V-A}\simeq (4.49\pm 1.18)\times 10^{-2}~.
\eeq
This result is in good agreement with the result obtained by using the
ALEPH/OPAL fitted mean value
for $\delta^{(6)}_{V/A}$:
\beq\label{aleph}
\delta^{(6)}_{V-A}\vert_{fit}\simeq (4.80\pm 0.29)\times 10^{-2}~.
\eeq
We shall use as a final result the average of these two determinations,
which coincides with
the most precise one in Eq. (\ref{aleph}). We shall also use the
result:
\beq\label{ratio78}
  \frac{\la{\cal
O}^{3/2}_7\ra}{\la {\cal O}^{3/2}_8\ra}\simeq \frac{1}{8.3}~\ga {\rm
resp.}~ \frac{3}{16}\dr~,
\eeq
where, for the first number we use the value of the ratio of $B^{3/2}_7/
B^{3/2}_8$ which is about $0.7\sim 0.8$ from e.g. lattice calculations
quoted in Table 1, and
the formulae in Eqs. (\ref{soft}) to (\ref{soft2}); for the second number
we use the vacuum saturation
for the four-quark vacuum condensates \cite{SVZ}. The result in Eq.
(\ref{ratio78}) is also comparable with the
estimate of \cite{DONO} from the sum rules given in Eq.(\ref{dispersive}).
Therefore, at the scale
$\mu=M_\tau$, Eqs. (\ref{o8}), (\ref{aleph}) and (\ref{ratio78}) lead, in
the $\overline{MS}$--scheme, to:
\beq\label{resO8}
\la{\cal O}^{3/2}_8\ra\ga M_\tau\dr\simeq -(0.94\pm 0.21)\times
10^{-3}~{\rm GeV}^6~,
\eeq
where the main errors come from the estimate of the unknown higher order
radiative corrections.
It is instructive to compare this result with the one using the vacuum
saturation assumption for the
four-quark condensate (see e.g. BNP):
\beq\label{O8}
\la{\cal O}^{3/2}_8\ra|_{v.s}\simeq -\frac{32}{18}\la \bar uu\ra^2\ga
M_\tau\dr\simeq -0.65\times
10^{-3}~\rm{GeV}^6~,
\eeq
which shows a $1\sigma$ violation of this assumption. This result is not
surprising as analogous violations have been obtained in other channels
\cite{SNB}. We have
used for the estimate of
$\la
\bar\psi\psi\ra$the value of $(m_u+m_d)(M_\tau)\simeq 10$ MeV \cite{SNL}
and the GMOR pion PCAC relation. However,
this violation  of the vacuum saturation is
not quite surprising, as a similar fact has also been observed in other
channels
\cite{SNB,ALEPH,OPAL}, though it also appears that the vacuum saturation
gives a quite
good approximate value of the ratio of the condensates
\cite{SNB,ALEPH,OPAL}. The result in Eq. (\ref{resO8})
is comparable with
the value $-(.98\pm 0.26)\times 10^{-3}~{\rm GeV}^6$ at $\mu$=2 GeV
$\approx M_\tau$ obtained by \cite{DONO}
using a DMO--like sum rule, but, as discussed previously, the DMO--like sum
rule result is very sensitive to
the value of $\mu$ if one fixes $t_c$ as in Eq. (\ref{tc}) according to the
criterion discussed above. Here,
the choice $\mu=M_\tau$ is well-defined, and then the result becomes more
accurate (as mentioned
previously our errors come mainly from the estimated unknown $\alpha_s^3$
term of the QCD series).
Using Eqs. (\ref{soft}) and (\ref{ratio78}), our previous result in Eq.
(\ref{resO8}) can be translated into the
prediction on the weak matrix elements in the chiral limit:
\beq\label{resq8}
 \la (\pi\pi)_{I=2}|{\cal Q}_{8}^{3/2}|K^0\ra(M^2_\tau)\simeq (2.58\pm
0.58)~{\rm
GeV}^3~\ga\frac{92.4}{f_\pi~[{\rm MeV}]}\dr^3~,
\eeq
normalized to $f_\pi$, which avoids the ambiguity on the real value of
$f_\pi$ to be used in a such expression.
Our result is higher by about a factor 2 than the quenched
lattice result \cite{ROME,MARTI}. A
resolution of this discrepancy can only be done after the inclusion of
chiral corrections in Eqs.
(\ref{soft}) to (\ref{soft2}), and after the uses of dynamical fermions on
the lattice. However, some
parts of the chiral corrections in the estimate of the vacuum condensates
are already included into the
QCD expression of the
$\tau$-decay rate and these corrections are negligibly small. We might
expect that chiral corrections,
 which are smooth functions of $m^2_\pi$ will not affect
strongly the relation in Eqs. (\ref{soft}) to (\ref{soft2}), though an
evaluation of their exact size is
mandatory.
Using the previous mean values of the light quark running masses
\cite{SNL}, we deduce in the
chiral limit and at the scale $M_\tau$:
\beq\label{resb8}
B^{3/2}_8(M^2_\tau)\simeq (1.70\pm 0.39)\ga\frac{m_s(M_\tau)~[{\rm
MeV}]}{119}\dr^2\ga\frac{92.4}{f_\pi~[{\rm
MeV}]}\dr^4~.
\eeq
 One should notice
that,  contrary to the $B$-factor, the result in Eq. (\ref{resq8}) is
independent to leading order on value of the light quark masses.
\section{NEW ALTERNATIVE SUM RULES FOR $\la{\cal O}^{3/2}_7\ra$}
\nin
Here, we shall attempt to present new sum rules for extracting ${\cal
O}^{3/2}_7$. In so doing, we work with the
renormalized $D=6$ condensate contributions to the difference of the vector
and axial-vector two-point
correlators. Using the expression given in the Appendix of BNP \cite{BNP},
one can deduce in the
$\overline{MS}$--scheme and in the chiral limit:
\bea
(-q^2)^3\Pi_{LR}(q^2)&=&4\pi\alpha_s(\mu)\aga \ga
1+\frac{119}{24}a_s(\mu)\dr\la{\cal
O}^{3/2}_8\ra+a_s(\mu)\la{\cal O}^{3/2}_7\ra\adr\nnb\\
&&-2\alpha_s^2(\mu)\ga\log {\frac{-q^2}{\mu^2}}\dr\ga -{\cal
O}^{3/2}_8+\frac{8}{3}{\cal O}^{3/2}_7\dr~,
\eea
which has the generic
form:
\beq
 q^6\Pi_{LR}(q^2) \sim A~\alpha_s(\mu)+B~\alpha_s^2(\mu)+C~\alpha_s^2(\mu)
\log{\frac{-q^2}{\mu^2}}~,
\eeq
where, we remind that $a_s \equiv \alpha_s/\pi$ is the renormalized
coupling and $A,~B,~C$ are constant numbers.
One can notice that by working with the different $q^2$--derivatives of
$(-q^2)^3\Pi_{V-A}(q^2)$, one can
eliminate the effects of the $A$ and $B$ terms, and derive the  Laplace sum
rule:
\beq\label{laplace}
\tau\int_0^{t_c} ds~ s^3~e^{-s\tau}\frac{1}{\pi}\rm{Im}\Pi_{LR}(s)\simeq
-2\alpha_s^2(\tau)(1-e^{-t_c\tau})\ga  -\la{\cal
O}^{3/2}_8\ra+\frac{8}{3}\la{\cal O}^{3/2}_7\ra\dr~,
\eeq
where we have transferred in the RHS the QCD continuum effect starting from
the threshold $t_c$.
Alternatively, we can also derive a
$\tau$-like sum rule \footnote{The coefficient has been checked using a
compilation of \cite{CHETY}.}:
\beq\label{taulike}
\int_0^{t_c} \frac{ds}{t_c}~s^3\ga 1-\frac{s}{t_c}\dr^n\frac{1}{\pi}{\rm
Im}\Pi_{LR}(s)\simeq
-2{\alpha_s^2(t_c)}\ga\frac{1}{n+1}\dr\ga  -\la{\cal
O}^{3/2}_8\ra+\frac{8}{3}\la{\cal
O}^{3/2}_7\ra\dr~.
\eeq
Formally, these sum rules are much better than the one proposed in
\cite{DONO,ENJL}, as the
leading $\mu$--dependence has disappeared after taking different
derivatives or after performing the Cauchy
integral. However, unlike the ones in \cite{DONO,ENJL}, one has gained one
power of $s$, which
renders the analysis more sensitive to the high-energy tail of the spectral
functions where the
data near the $\tau$--mass are quite bad \cite{ALEPH,OPAL}. One should also
notice that here the
RHS starts at order $\alpha_s^2$, which means that chiral corrections can
affect dangerously the
RHS of the sum rules, and can compete with the dimension-six condensate
contributions. Using the
present ALEPH/OPAL data on the
$V-A$ spectral functions, and a traditional
$\tau$--stability analysis of the Laplace sum rule, we realize that the
$\tau$--stability  is reached at exceptional
large $\tau$--values of about $(1.8\sim 3)$ GeV$^{-2}\approx M^{-2}_\rho$,
where the
OPE can already break down. In
the case of the
$\tau$--like sum rule, one finds that for a given value of $t_c$, one has a
$n$--stability where the value of $n$
increases with $t_c$ ($n\geq 2.5$). However, at a such value of $n$, one
needs more and more
information (by duality) on the non-perturbative contributions to the sum
rules.
From our analysis and using the value of $\la{\cal
O}^{3/2}_8\ra$ obtained in Eq. (\ref{resO8}), we obtain the conservative range:
\beq
-2\times 10^{-3}\leq \la{\cal
O}^{3/2}_7\ra \leq 10^{-2}~{\rm GeV}^6,
\eeq
where the lower (resp. higher) value corresponds to the first (resp.
second) set of $t_c$--values given in Eq.
(\ref{weinberg}). Therefore, we conclude that in order to extract more reliable
informations on the previous sum rules, one needs to have much more
informations on these sum rules both experimentally
and theoretically. In particular, these sum rules can be  more useful when
accurate data near the
$\tau$--mass is available and all chiral symmetry breaking terms are
included both in the RHS of the sum
rule and in the derivation of the relation between the kaon matrix elements
and the vacuum condensates.
\section{$I=0$ SCALAR MESON CONTRIBUTION TO  $\la (\pi^+\pi^-)_{I=0}|{\cal
Q}_6^{1/2}|K^0_S\ra$}
\nin
We study the effect of a direct production of a $I=0$ scalar meson
intermediate state, which we shall denote by $S_2\equiv (\bar uu +\bar
dd)$, to the $K^0_S\rar(\pi^+\pi^-)_{I=0}$
decay process. Before doing our analysis, let us present the status of the
scalar meson spectrum below 1 GeV.
\subsection{A short review on the scalar meson spectrum below 1 GeV}
\nin
A much more complete review and analysis of the complex structure of scalar
mesons spectra is given in
\cite{SNG}. Here, we shall be concerned with the spectrum of light scalar
mesons below 1 GeV:
\beq
a_0(980), ~~~~~ f_0(975),~~~~~\sigma(400\sim 1200)~,
\eeq
as quoted by PDG \cite{PDG}. We associate the isovector state $a_0(980)$ to
the divergence of the vector
current:
\beq
\partial_\mu V^\mu_{\bar ud}=(m_u-m_d) \bar u (i) d~,
\eeq
which is natural from the point of view of chiral symmetry
\cite{PNR,SNB,SNG,BRAMON} and in the
construction of an effective chiral Lagrangian including resonances
\cite{SCALEFF}. In this scheme the
$K^*_0(1.43)$ \cite{PDG} is the $\bar su$ partner of the $a_0$. The
predicted hadronic and two-photon
widths of the
$a_0$ using vertex sum rules
\cite{SNB,BRAMON,SNG} are in good agreement with present  data
\cite{MONT,PDG}. These data indeed confirm the
$\bar qq$ nature of the
$a_0$ and disfavour some other exotic interpretations, which can further be
tested through measurements of the
$\phi$ radiative decay at Daphne.
\\ In our analysis of the $I=0$ isoscalar channel, we consider the trace of
the energy momentum tensor:
\beq
\theta^\mu_\mu= \frac{1}{4}\beta(\alpha_s) G^2 +\ga
1+\gamma_m(\alpha_s)\dr\sum_{u,d,s} m_i\bar\psi_i\psi_i~,
\eeq
where $\beta$ and $\gamma_m$ are the $\beta$ function and mass anomalous
dimension. We shall consider the bare
states before mixing:
\begin{itemize}
\item The
meson
$S_2$ is associated to $(\bar uu+\bar dd)$, which is degenerate to the
$a_0$ because of the good $SU(2)$
symmetry.
\item The $S_3\equiv\bar
ss$ state is above 1 GeV due to
$SU(3)$ breaking
\cite{SNB,SNG}, and will not give significant effects in our analysis.
\item The meson
$\sigma_B$ is a gluonium state, which has been needed for solving
\cite{VENEZIA} the inconsistencies between the substracted
\cite{NSVZ} and unsubstracted \cite{SNGLUE} gluonium sum rules
\footnote{The resolution of such
inconsistencies has been also improved recently by the inclusion of the new
$1/q^2$-term induced by the
tachyonic gluon mass in the OPE \cite{CNZ}.}. The $\sigma_B$ mass is
expected to be around $0.7\sim 1$ GeV,
but it will not play a significant r\^ole in this analysis, as the
gluonium-quarkonium mixing in the propagator
(mass mixing) via the off-diagonal two-point correlator is small
\cite{MENES,SNB}.
\end{itemize}
The
separation of the quark and gluon components of the current is allowed by
renormalization group invariance (RGI) as
$m\bar\psi\psi$ is RGI, while $\alpha_s G^2$ only mix with $m\bar\psi\psi$
to higher order in $\alpha_s$
\cite{TARRACH}.   The hadronic couplings, decay constants and masses of
these mesons have been estimated
using vertex sum rules
\cite{BRAMON,MENES,SNG}, low-energy theorems \cite{VENEZIA,SNB,SNG}, and/or
some $SU(3)$ symmetry relations among
the meson wave functions \cite{BRAMON}. It comes out that:
\begin{itemize}
\item The
$\sigma_B$ couples strongly and universally to pairs of Goldstone bosons
(large violation of the OZI rule)
\footnote{This ``decay mixing" which occurs via 3-point function should not
be confused with the ``mass mixing"
via an off-diagonal 2-point function. There is not a contradiction between
a large ``decay mixing" and a small
``mass mixing".}
\cite{VENEZIA,SNB,SNG}, which invalidates the lattice results in the
quenched approximation \footnote{The gluonium
mass obtained in the quenched approximation of about 1.5 GeV can be
indentified with the one of about 1.6 GeV
obtained from the unsubtracted sum rule \cite{SNGLUE,SNG}, which is shown
\cite{VENEZIA,SNG} to couple weakly to Goldstone pairs but strongly to glue
rich $U(1)_A$ states like
$\eta'$-$\eta'$, through mixing to $\eta'$-$\eta$, and to $4\pi$ through
$\sigma_B$-$\sigma_B$ pairs.}.
\item The
$S_2$ is relatively narrow with a width of about 120 MeV and couples almost
equally to $\pi\pi$ and $\bar KK$.
\item The $I=0$ scalar spectrum below 1 GeV, i.e., the observed wide
$\sigma$--meson seen below 1 GeV
\cite{PDG,MENESSIER} and the narrow
$f_0(980)$\cite{PDG} states, is expected in our approach to come from a
maximal mixing between the
$S_2\equiv (\bar uu+\bar dd)$ and the gluonium $\sigma_B$ bare states.
\item Recent data favour such a maximal gluonium-quarkonium
mixing scheme \cite{MONT}, together with the $\bar qq$ nature of the
isovector $a_0(980)$ state, though further
refined tests are still needed.
\end{itemize}
\subsection{Parameters of the $S_2\equiv (\bar uu+\bar dd)$ scalar meson}
\nin
In the following, we shall give the values of the decay constant and
couplings of the $S_2$ which is the relevant
particle in the present analysis.
\begin{itemize}
\item Its
decay constant has been fixed using Laplace sum rule for the associated
two-point correlator. At the minimum of the
sum rule variable
$\tau_0$
\footnote{The inclusion of the effect of tachyonic gluon increases the
value of $\tau$ from 0.5 GeV$^{-2}$
\cite{SNG,SNB} to 1 GeV$^{-2}$ \cite{CNZ}, improving the duality between
the resonance and the QCD sides of
the sum rules, but does not almost affect the result, like in the case of
the pion sum rule.}, and at the inflexion
point of its change versus the QCD continuum threshold of about 2.6
GeV$^2$, it has the value
\cite{SNB,SNG}:
\beq
{f_S}/{(m_u+m_d)(\tau_0)}\simeq (0.32\pm 0.08)~,
\eeq
where, presented in this way, the number in the RHS is not sensitive to the
change of quark mass values, but has
an anomalous dimension, and it runs like the inverse of the quark mass. It
is normalized as:
\beq
\frac{1}{\sqrt{2}}{(m_u+m_d)}\la 0\vert \bar uu+\bar dd\vert
S_2\ra=\sqrt{2}{f_S M^2_S}~.
\eeq
One should notice that using the value of $(m_u+m_d)(\tau_0)$ given in
\cite{SNL}, the value of $f_S$ is
about 2 MeV, which is much smaller than $f_\pi$, and which invalidates the
estimate $f_S\approx f_\pi$
often proposed in the literature. Instead, it is the quantity
$M^2_Sf_S\approx m^2_\pi f_\pi$, which is almost
constant.
\item The $S_2$ hadronic coupling to $\pi\pi$
has been fixed using leading order results from vertex sum rules. It reads
\cite{BRAMON,MENES,SNB,SNG}:
\beq
g_{S\pi^+\pi^-}\approx \frac{16\pi^3}{3\sqrt{3}}\la \bar uu\ra \tau_0 ~{\rm
exp}\ga M^2_S\frac{\tau_0}{2}\dr\simeq
2.5 ~{\rm GeV},
\eeq
corresponding to $\tau_0\simeq 1$ GeV$^{-2}$. It leads to the decay width
$\Gamma(S_2\rar\pi^+\pi^-)\simeq 120$
MeV, with the normalization:
\beq
\Gamma\ga S_2\rar\pi^+\pi^-\dr=\frac{\vert g_{S\pi^+\pi^-}\vert^2}{16\pi
M_S}\ga
1-\frac{4m^2_\pi}{M^2_S}\dr^{1/2}~.
\eeq
This result is in good agrement with the
one obtained from the
$a_0\eta\pi$ coupling, by using  $SU(3)$
symmetry for the meson wave functions \cite{BRAMON}:
\beq
g_{S\pi^+\pi^-}\simeq \sqrt{\frac{3}{2}}g_{a_0\eta\pi}\simeq (2.50\pm
0.15)~{\rm GeV},
\eeq
where we have used the peak data: $\Gamma (a_0\rar\eta\pi)\simeq (57\pm 7)$
MeV \cite{PDG}. One should notice that
this coupling does not vanish in the chiral limit because it behaves like
$\la \bar uu\ra$, and, up to
$SU(3)$-breakings, one expects an universal coupling of the $S_2$ to
Goldstone boson pairs.
\end{itemize}
We shall use these parameters as inputs in the following analysis.
\subsection{The {$S_2\equiv (\bar uu+\bar dd)$} scalar meson contribution to
{ $K_S\rar (\pi^+\pi^-)_I=0$}
decay}
\nin
\begin{itemize}
\item
We shall work with on-shell kaon, such that the tadpole diagram will not
contribute in our analysis \cite{GAVELA}. We can write, in the chiral limit:
$m_u=m_d=0$ and
$\la
\bar ss\ra=\la \bar dd\ra$ \footnote{We follow the notations and
conventions of \cite{VSZ}.}:
\bea\label{matrix}
\la {\cal Q}^{1/2}_{6}\ra_{2\pi}\equiv\la (\pi^+\pi^-)_{I=0}|{\cal
Q}_6^{1/2}|K^0\ra&\simeq&-\Big{[}2\la\pi^+|\bar
u\gamma_5 d|0\ra\la\pi^-|\bar su|K^0\ra+\nnb\\
&&\la \pi^+\pi^-|\bar dd+\bar uu |0\ra\la 0|\bar s\gamma_5
d|K^0\ra~\Big{]}.
\eea
For convenience, we shall evaluate the matrix elements at the scale
$\mu=m_c$ because the Wilson
coefficients are also given at this scale \cite{BURAS}.
We shall use the
value of $m_c$ from combined sum rule analysis of the charmonium and $D$-meson
\cite{SNH} and the QCD spectral sum rule average of the light quark masses
quoted in \cite{SNL}. They read
at the scale $\mu=m_c$:
\bea
m_c(m_c)&\simeq& (1.20\pm 0.05)~{\rm GeV},~~
m_s(m_c)\simeq (147\pm 15)~ {\rm MeV},\nnb\\
m_d(m_c)&\simeq& (8\pm 1)~{\rm MeV}~~m_u(m_c)\simeq(4.4\pm 0.5)~{\rm MeV}~.
\eea
\item The first term of the weak matrix element is well-known, and can be
related to the
$K\rar
\pi~ l\nu_l$ semi-leptonic form factors (see e.g. \cite{VSZ}):
\beq
\la\pi^-|\bar su|K^0\ra=\Big{[}f_+(M^2_K-m^2_\pi)+f_-m^2_\pi\Big{]}/(m_s-m_u)~,
\eeq
with, to leading order in the chiral symmetry breaking terms: $f_+\approx
1$ and $f_-\approx 0$. It leads to:
\bea\label{fplus}
\la\pi^+|\bar u\gamma_5
d|0\ra\la\pi^-|\bar su|K^0\ra(m_c)&\simeq& \frac{\sqrt{2}f_\pi
m^2_\pi}{(m_d+m_u)}
\frac{m^2_K-m^2_\pi}{(m_s-m_u)}\nnb\\
&\simeq& (0.323\pm 0.032)\ga \frac{142.6}{(m_s-m_u)~[{\rm
MeV}]}\dr^2{\rm GeV}^3,
\eea
where $m_i$ are the running quark masses evaluated at the scale $m_c$.
Chiral corrections to
these terms are known to be about 10\% in the literature (see e.g.
\cite{SIGMAN}), which
have been included into the error estimate.
\item For the second term, we assume that it is dominated by the direct
production of the $S_2$--scalar
meson in the $s$--channel for an on-shell kaon $p^2=m^2_K$. Therefore, it
can be decomposed
as:
\beq\label{sigma}
\la \pi^+\pi^-|\bar dd+\bar uu |0\ra =\la \pi^+\pi^-|S_2\ra\la S_2|\bar
dd+\bar uu |0\ra\equiv
\frac{g_{S\pi^+\pi^-}}{\ga p^2-M^2_S\dr}\frac{2f_S}{(m_u+m_d)}M_S^2~.
\eeq
With the values of the parameters given
previously, we conclude
that the scalar meson contribution is:
\beq\label{s2}
\la \pi^+\pi^-|\bar dd+\bar uu |0\ra\la 0|\bar s\gamma_5
d|K^0\ra(m_c)\simeq \pm{(0.53\pm 0.13)~\ga \frac{155}{(m_s+m_d)~[{\rm
MeV}]}\dr{\rm GeV}^3}~,
\eeq
where we take into account the fact that QSSR cannot fix the signs of the
$S_2$ coupling and decay
constant, which will be fixed later on from chiral constraints on the weak
amplitude \cite{REFEREE}.
The error in this determination comes mainly from the one of decay constant
$f_S$.
\item For on-shell kaon, and neglecting, to a first approximation,
$m_{u,d}$ (resp.
$m_\pi^2)$ versus $m_s$ (resp. $m_K^2)$, we deduce from Eqs.
(\ref{matrix}) to (\ref{sigma}), the approximate relation:
\beq
\la {\cal Q}^{1/2}_{6}\ra_{2\pi} (m_c)\approx -\frac{2\sqrt{2}f_\pi
m^2_\pi}{(m_d+m_u)}
\frac{m^2_K}{m_s}\Bigg{[}
-1+\frac{f_K}{f_\pi}\frac{g_{S\pi^+\pi^-}f_S}{m_\pi^2}{\ga
1-\frac{M^2_K}{M^2_S}
\dr}^{-1}\Bigg{]}~,
\eeq
which we can consider as an updated version of the expression given by VSZ
in \cite{VSZ}.
satisfying the double  chiral constraint conditions (vanishing of the
amplitude when $M_K^2 \rar 0$
and $f_K=f_\pi$) \cite{GAVELA}, which are recovered if
\beq
\frac{g_{S\pi^+\pi^-}f_S}{m_\pi^2}{\ga
1-\frac{M^2_K}{M^2_S}
\dr}^{-1}\approx 1~,
\eeq
This value is obtained within the errors from the values of the
$S_2$-parameters given previously.
\item
For the numerics, we shall use more precise values of the different
parameters by keeping corrections to
order
$m_\pi^2$ and $m_{u,d}$. Therefore, taking
Eqs. (\ref{fplus}) and (\ref{s2}) into Eq. (\ref{matrix}), one obtains the
final
value of the weak matrix element:
\bea\label{resq6}
\la {\cal Q}^{1/2}_{6}\ra_{2\pi} (m_c)\simeq& -& \ga
\frac{142.6}{(m_s-m_u)~[{\rm MeV}]}\dr^2
\times\Bigg{[} \ga 0.65\pm 0.09\dr\nnb\\
 &-&{(0.53\pm 0.13)} \ga \frac{(m_s-m_u)~[{\rm
MeV}]}{142.6}\dr\Bigg{]} ~{\rm GeV}^3.
\eea
Using the relation \cite{BURAS} \footnote{We shall use the usual
parametrization in terms of $m_s^2$,
but it can be misleading in view of the $m_s$-dependence of our results in
Eqs. (\ref{fplus}) and (\ref{s2}).}:
\beq
\la {\cal Q}^{1/2}_{6}\ra_{2\pi} (m_c)\simeq -4\sqrt{\frac{3}{2}}\ga
\frac{m^2_K}{m_s+m_d}\dr^2\sqrt{2}\ga
f_K-f_\pi\dr B^{1/2}_6(m_c)~,
\eeq
where $f_K\simeq 1.22 f_\pi$, the previous result can be translated
into:
\beq\label{resb6}
B_6^{1/2}(m_c)\simeq 3.7 \ga \frac{m_s+m_d}{m_s-m_u}\dr^2 \Bigg{[} \ga 0.65\pm
0.09\dr-{(0.53\pm 0.13)} \ga
\frac{(m_s-m_u)~[{\rm MeV}]}{142.6}\dr\Bigg{]}
\eeq
Evaluating the running quark masses at 2 GeV, with the values given
previously, one deduces:
\bea\label{resb6a}
B_6^{1/2}(2)&\simeq &(1.0\pm 0.4) ~{\rm for ~}m_s(2)=119~{\rm MeV},\nnb\\
&\leq&(1.5\pm 0.4) ~{\rm for ~}m_s(2)\geq 90~{\rm MeV}~.
\eea
The errors added quadratically have been relatively enhanced by the partial
cancellations of the two
contributions.
\end{itemize}
\section{COMPARISON OF OUR RESULTS WITH SOME OTHER PREDICTIONS}
In this section, we shall compare our values of $B^{3/2}_{7,8}$ and
$B^{1/2}_6$ with the results in Table
1.
\subsection{Value of $B^{3/2}_7$}
\nin
\begin{itemize}
\item Our value  in Eq. (\ref{b78}):
\beq
B^{3/2}_7(\mu=2~{\rm GeV})\simeq (0.7\pm 0.2)\ga\frac{m_s(2)~[{\rm
MeV}]}{119}\dr^2\ga\frac{92.4}{f_\pi~[{\rm MeV}]}\dr^4~,
\eeq
 comes from a re-analysis of the DMO--like sum rule used in \cite{DONO}, and in
\cite{ENJL} within a large $N_c$ expansion and a lowest meson dominance
(LMD). One can notice in Table 1 a quite good
agreement  between the results from different approaches. However, due to the strong
$\mu$-dependence of the result, one should be careful when giving its value.
\item Our analysis from the Laplace and $\tau$--like sum rules are
unfortunately
inconclusive using present
$\tau$-decay data and present theoretical approximation (chiral limit).
\end{itemize}
\subsection{Value of $B^{3/2}_8$}
\nin
Our result in Eq. (\ref{resb8}):
\beq
B^{3/2}_8(M^2_\tau)\simeq (1.7\pm 0.4)\ga\frac{m_s(M_\tau)~[{\rm
MeV}]}{119}\dr^2\ga\frac{92.4}{f_\pi~[{\rm
MeV}]}\dr^4~,
\eeq
comes from the analysis of the total $\tau$ hadronic width $R_{\tau,V-A}$.
\begin{itemize}
\item
The closest comparison to be made is the one with \cite{DONO} where soft
pion and kaon reductions together
with DMO sum rules have been used. We have stressed that in order to obtain
our value, we have not introduced
any new sum rule but took advantage of the existing measurement of the
vector and axial-vector
components of the $\tau$--total width and the measured values of the
corresponding $D=6$ vacuum condensates.
\item Our result agrees numerically within the errors with the one of
\cite{DONO}, but we have also shown that the
DMO sum rules lead to an inaccurate value due mainly to the bad convergence
of the QCD series.
\item Our result includes NLO corrections, an estimate of the NNLO terms
and the effect of a tachyonic gluon mass
which phenomenologically takes into account the resummation of the QCD
asymptotic series.
\item Our
result is in the range given by some linear \cite{SIGMAL} and non-linear
\cite{SIGMAN} $\sigma$ models, but
is 1 to 2$\sigma$ higher than the largest values obtained from the present
lattice \cite{MARTI,GUPTA}, large
$N_c$
\cite{HAMBYE}, chiral quark model \cite{FAB} and the one including final
state interactions \cite{PALLANTE}.
Though the agreements with the results from the $\sigma$ models are
interesting, it is not clear to us how to
connect the two approaches, as in these models, an $I=0$ scalar resonance
has been introduced with the
parameters of the observed $\sigma$--meson which we expect \cite{VENEZIA,SNB,SNG} to
have a large gluon component in its wave function.
The difference with the result from final state interactions
\cite{PALLANTE} is more rewarding, and needs
a much better understanding in connection to the comments raised in
\cite{BURAS2}.
\end{itemize}
\subsection{Value of $B^{1/2}_6$}
\nin
Assuming a dominance of the $I=0$ $S_2\equiv (\bar uu+\bar dd)$ scalar
meson contribution, through the operator
${\cal Q}^{1/2}_6$, to
the { $K_S\rar (\pi^+\pi^-)_I=0$} decay amplitude, we have obtained:
\beq\label{resb6}
B_6^{1/2}(m_c)\simeq 3.7 \ga \frac{m_s+m_d}{m_s-m_u}\dr^2 \Bigg{[} \ga 0.65\pm
0.09\dr-{(0.53\pm 0.13)} \ga
\frac{(m_s-m_u)~[{\rm MeV}]}{142.6}\dr\Bigg{]}~,
\eeq
leading to:
\bea\label{resb66}
B_6^{1/2}(2)&\simeq &(1.0\pm 0.4) ~~~{\rm for ~~~}m_s(2)=119~{\rm MeV},\nnb\\
&\leq&(1.5\pm 0.4) ~~~{\rm for ~~~}m_s(2)\geq 90~{\rm MeV}~,
\eea
which we give in Table 1.
\begin{itemize}
\item Contrary to {\it conventional} approaches
\cite{RAFAEL,BURAS,MARTI,VSZ}, which do not consider the effect of a scalar
meson, our result shows that the $\bar qq$ component of the scalar meson
tends to brings the value of
$B^{1/2}_6$  to the one of the leading $1/N_c$
expectation.
\item In our analysis,
the mass of the
$S_2$ is fixed from
$SU(2)$ symmetry arguments, and obtained from the two-point function sum
rule to be about the one of the
$a_0(980)$, which is relatively high compared with the kaon mass.
Therefore, the main contribution
observed here comes from the values of the $S_2$-coupling to
$\pi\pi$ and of its decay constant. However, for a more definite
conclusion, it is important
to look for the effect of the gluon component of the $\sigma$, which can
eventually give a sizeable contribution in the amplitude through a new
operator other than the one discussed here.
This new feature may clarify the observed enhancement from a direct final
state interactions (FSI) analysis  of
the amplitude. We plan to come back to this point in a future work.
\item Present lattice results \cite{MARTI,GUPTA} are still unreliable
\cite{KILCUP}, as the NLO QCD
corrections at the matching scale between the lattice and continuum results
are huge. Measuring the effects
of the scalar meson on the lattice seems to be difficult due to the
propagator \cite{MARTINELLI}.
\item Due to the partial concellations of the two contributions in the weak
amplitude, taking into account the
alone effect of the $\bar qq$ component of the isoscalar meson is not
sufficient for explaining the large
enhancement obtained from some other approaches quoted in Table 1 which we
list below:
\begin{itemize}
\item Some incomplete large $N_c$ result \cite{HAMBYE} including higher
order corrections ${\cal O}(p^2/N_c)$ in
the chiral limit.
\item A version of the ENJL-model \cite{RAFAEL} with an intermediate vector
bosons \cite{PRADES} and the
chiral quark model \cite{FAB} where both models are based on the $1/N_c$
expansion. However, for the chiral quark
model the predictions correspond to a lower value of the scale
$\mu$. A clear connection with these results with the $\overline{MS}$
scheme as well as the relation of the
parameters used there with lattice and QSSR calculations is needed.
\item Final state interactions \cite{PALLANTE} where there is still a
debate for fixing the slope of the amplitude
\cite{BURAS2}.
\item Enhancements due to the isoscalar meson have also been found from the
$\sigma$--model
approaches
\cite{SIGMAN,SIGMAL}, which is mainly due to the small value of the
$\sigma$-mass used in the
$\sigma$-propagator appearing in Eq. (\ref{sigma}). However, the
uncertainties come from the fact that, in the
linear sigma models
\cite{SIGMAL} the Lagrangian is not unique, while in the non-linear
$\sigma$ models \cite{SIGMAN}, the
$\sigma$ mass is a free parameter which is usually identified with the
observed wide $\sigma$-meson having
a mass in the range ($0.4\sim 1.2$) GeV and a width of about $
(600\sim 1000)$ MeV \cite{PDG,MENESSIER}. On the contrary, in the
present work, for the reasons previously explained, the scalar meson entering
into the analysis is not the observed $\sigma$ where, within our framework,
the $\sigma$ comes from a maximal
(decay) mixing between a $S_2$ ($\bar qq$) and gluonium ($\sigma_B$)
states. Indeed, as explained in the
previous sections, only the
$\bar qq$ component (the hypothetical
$S_2$ state) of the $\sigma$  is relevant for the present operator
\footnote{The effect of the gluon component
(called
$\sigma_B$) through the scalar propagator is negligble (mass mixing)
\cite{MENES,SNB,SNG} as it comes from
the off-diagonal quark-gluon two-point function. This fact does not
contradict the large gluonium decay into
$\pi\pi$ (decay mixing) which comes from a vertex function
\cite{VENEZIA,SNB,SNG}.}. However, some advanced versions of the effective
Lagrangian approach, which can
separate explicitly the $\bar qq$ from the gluon component of the scalar
meson, are needed for the present problem
\footnote{Some attempts to introduce the $I=0$ scalar meson within the
effective lagrangian framework exist in the
literature \cite{SCALEFF,SIGMAN,SIGMAL,NAGY}.}.
\end{itemize}
\item A clever explanation showing the connections of the different results
reviewed here in Table 1 in
order to have an unified explanation of these different determinations is
still needed.
\end{itemize}
\subsection{VALUES OF $B^{1/2}_6/B^{3/2}_8$ }
\nin
Using our previous determinations of $B_8^{3/2}(M_\tau)$ in Eq.
(\ref{resb8}) and $B_6^{1/2}(2)$
in Eq. (\ref{resb6a}), and the previous value of $m_s(2)$, one can
deduce the ratio:
\beq\label{r68}
{\cal R}_{68}\equiv\frac{B_6^{1/2}}{ B_8^{3/2}}\simeq 0.6\pm 0.3~,
\eeq
and their combination:
\bea\label{comb}
{\cal B}_{68}\equiv B_6^{3/2}-0.48B_8^{3/2}\simeq (0.3\pm 0.4)~~{\rm for}~~m_s(2)=119~{\rm
MeV}~,
\eea
where we have added the errors quadratically. \\
Instead, using the lower bound $m_s(2)\geq
90~{\rm MeV}$ reported in \cite{SNL} into the expressions of $B_8^{3/2}(M_\tau)$ in Eq.
(\ref{resb8}) and $B_6^{1/2}(2)$, one can deduce the {\it conservative} upper bound:
\bea\label{comb2}
{\cal B}_{68}\equiv B_6^{3/2}-0.48B_8^{3/2}\leq (1.0\pm 0.4) ~~{\rm for}~~m_s(2)\geq 90~{\rm
MeV}~,
\eea
where again we have added the errors quadratically.
\section{VALUE AND UPPER BOUND OF $\epsilon'/\epsilon$}
\nin
\begin{itemize}
\item
The estimated value in Eq. (\ref{comb}) does not satisfy the constraint required in Eq. (\ref{const}) for
explaining the present data on the
$CP$-violation ratio
$\epsilon'/\epsilon$ given in Eq. (\ref{exp}). It leads, for $m_s(2)=119$ MeV, to the prediction:
\beq\label{resepsilon2}
\frac{\epsilon'}{\epsilon}\simeq (4\pm 5)\times 10^{-4}~.
\eeq
The failure for reproducing the data
may indicate the need for other contributions than the alone
$\bar qq$ scalar meson $S_2$ (not the observed
$\sigma$)-meson for explaining, within the standard model (SM), these data.
Among others, a
much better understanding of the effects of the gluonium (expected large
component of the $\sigma$-meson
\cite{VENEZIA,SNG,BRAMON}) in the
amplitude, through presumably a new operator needs to be considered. 
\item If we use instead the {\it conservative} upper bound for
${\cal B}_{68}$ in Eq. (\ref{comb2}) corresponding to the lower
bound for the strange quark mass $m_s(2)\geq$ 90 MeV, we can deduce the bound
\footnote{The real value of $f_\pi$ (92.4 or 87 MeV) used in the chiral limit expression of
$B^{3/2}_8$ does not affect significantly the result.}:
\beq\label{resepsilon}
\frac{\epsilon'}{\epsilon}\leq (22\pm 9)\times 10^{-4}~.
\eeq
The errors come mainly from ${\cal B}_{68}$ (40\%) and Im
$\lambda_t$ (10.5\%), which we have added quadratically. In $B_6^{1/2}$,  the large error is due to the partial
cancellation of the contributions from the semi-leptonic form factors and the $S_2$ resonance. 
This bound agrees within the errors with the data on the $CP$-violation ratio
$\epsilon'/\epsilon$ given in Eq. (\ref{exp}). 
\end{itemize}
\section{COMMENTS ON THE $\Delta I=1/2$ RULE}
\nin
However, unlike Ref. \cite{SIGMAL}, we do not expect that our
result will affect significantly the $CP$--conserving $\Delta I=1/2$ rule
process. Indeed, according to the
analysis in \cite{RAFAEL,BURAS}, the amplitude Re$A_0$ of this process is
dominated by the pure $I=1/2$
combination:
\beq
Q_- \equiv Q_2-Q_1~,
\eeq
where its Wilson coefficient is relatively enhanced compared with the ones
of $Q_+ \equiv Q_2+Q_1$ and $Q_6$.
Moreover, the  one of $Q_6$, where the $S_2\equiv (\bar uu+\bar dd)$ can
contribute, can even be zero at the subtraction point
$\mu=m_c$ for a given renormalization scheme (so-called HV-scheme),
but still remains negligible at larger values of $\mu$ where the
perturbative calculations of these Wilson
coefficients can be trusted. Instead, octet scalar may play a rule in this
process as has been emphasized in the
first study of this process on the lattice \cite{MARTIN}. We
plan to analyze carefully this process in a future work.
\section{SUMMARY}
\begin{itemize}
\item We have used the recent ALEPH/OPAL data on the $V-A$ hadronic
spectral functions from $\tau$-decays for
fixing the matching scale separating the low and high-energy regions
(continuum threshold), at which the first
and second Weinberg sum rules should be realized in the chiral limit [Eq.
(\ref{tc})]. We have used this
information for predicting and for testing the accuracy of the low-energy
constants $m_{\pi^+}-m_{\pi^0}$
[Eq. (\ref{deltam})] and ${ L}_{10}$ [Eq. (\ref{l10})], and the electroweak
kaon penguin matrix elements $\la
(\pi\pi)_{I=2}|{\cal Q}_{8,7}^{3/2}|K^0\ra$ [Eqs. (\ref{dono78}) and
(\ref{b78})] obtained from DMO--like sum
rules in the chiral limit.
\item We have estimated the value of the  weak matrix element $\la
(\pi\pi)_{I=2}|{\cal Q}^{3/2}_{8}|K^0\ra$
using the measured $V/A$ $\tau$--decay rate and the experimentally fitted
value of the dimension six-operators,
without introducing any additionnal sum rules. Our results, in Eqs.
(\ref{resq8}) and (\ref{resb8}), indicate a
deviation from the vacuum saturation of the four-quark condensates, where
analogous violations have been
already found from the analysis of other channels \cite{ALEPH,OPAL,SNB}.
\item We have introduced some alternative new sum rules in Eqs
(\ref{laplace}) and (\ref{taulike}) in order to
estimate $\la (\pi\pi)_{I=2}|{\cal
Q}^{3/2}_{7}|K^0\ra$, which require improved data in the
region near the $\tau$--lepton mass and more theoretical inputs in order to
be useful. At present, the result from
the DMO--like sum rule obtained in Eq. (\ref{b78}) is more meaningful.
\item We remind that the results for $\la
(\pi\pi)_{I=2}|{\cal Q}_{8,7}^{3/2}|K^0\ra$ matrix elements have been
obtained in the chiral limit as we have
taken advantage of the soft pion and kaon reductions techniques in order to
express them in terms of the vacuum
condensates. Main improvements of our results need the inclusion of these
chiral corrections.
\item In the last part of the paper, we have analyzed the effect of the
$S_2\equiv(\bar uu+\bar dd$) component of
the
$I=0$ scalar meson into the
$\la (\pi\pi)_{I=0}|{\cal Q}^{1/2}_{6}|K^0\ra$ matrix element. We found
that its main contribution is due to the
values of its decay constant and coupling to $\pi\pi$ giving the
predictions in Eqs. (\ref{resq6}) and
(\ref{resb6}), but not on the enhancement due to its mass in the propagator.
\item
We have expressed our predictions on the matrix elements in terms of the
$B$--parameters, with the absolute values
given in Eqs. (\ref{resb8}), (\ref{resb6}) and rescaled at 2 GeV in Table
1, where we have used the value
$m_s(2$ GeV)$\simeq (119\pm 12)$ MeV \cite{SNL}. These results are compared
with the ones from different
approaches. The ratio of these $B$--parameters is also given in Eq.
(\ref{r68}) and compared with the existing values.
Their combination given in Eq. (\ref{comb}) is also compared with the
constraint in Eq. (\ref{const}).
\item
We have used our previous determinations of the penguin matrix elements in
order to predict the value of the
$CP$--violating ratio
$\epsilon'/\epsilon$. Our conservative upper bound in Eq. (\ref{resepsilon}) corresponding to $m_s(2)\geq$
90 MeV, agrees
with the present day experiments given in Eq. (\ref{exp}). However, our estimate in Eq. (\ref{resepsilon2})
corresponding to $m_s(2)\simeq$
119 MeV fails to explain the data, which is mainly due to the partial cancellation of the contributions implied
by the chiral constraints governing the
$\la (\pi\pi)_{I=0}|{\cal Q}^{1/2}_{6}|K^0\ra$ matrix element. This failure may not be quite surprising as
the observed $\sigma$-meson is expected to have a large gluonium admixture
responsible for its large $\pi\pi$ width \cite{VENEZIA,SNG,BRAMON},
which can manifest in the $K\rar\pi\pi$ amplitude through an eventual new
operator not considered until now, but most probably along the line of dimension-8 operators discussed
recently \cite{DONO3}. Further study of the gluonium effect is therefore mandatory before an
eventual consideration of
possible effects due to new physics.
\end{itemize}
\section*{ACKNOWLEDGEMENTS}
\nin
It is a pleasure to thank the KEK-theory group and Kaoru Hagiwara for their
hospitality. Stimulating discussions with Guido Martinelli in a preliminary
stage of
this work are acknowledged. The original form of the manuscript has
been improved thanks to the different questions and discussions during the
```Japan Tour"
at the universities of Ochanomizu (Tokyo),
Hiroshima, Kanazawa, Tsukuba, Sendai and
Kyoto, and at the Riken Institute (Tokyo); during the visit at the KIAS Institute (Seoul),
during the QCD 00 conference (Montpellier), and thanks to
the referee's report.
\vfill\eject

\end{document}